\documentclass[
    ,final            % use final for the camera ready runs
  ]
  {aipproc}

\layoutstyle{6x9}

\begin{document}

\title{Investigation of $K^0\Sigma^+$ photoproduction with the CBELSA/TAPS experiment}


\author{T.C.~Jude, for the CBELSA/TAPS and BGO-OD collaborations}{
  address={Universit\"at Bonn, Bonn, Germany}
}

\begin{abstract}

$\gamma (p,K^0)\Sigma^+$ cross section was measured at the CBELSA/TAPS experiment at the ELSA electron stretcher facility.  A cusp-like structure was apparent near the $K^*$ production threshold, where the cross section reduced by a factor of four in the forward direction.  It is speculated that t-channel $K^*$ exchange, and meson-hyperon dynamically generated states may have large contributions to the cross section.  The motivation for the extraction of the beam-target double polarisation observable, $E$, is discussed, and the new experiment, BGO-OD at the ELSA facility is introduced.

\end{abstract}

\maketitle

\section{Introduction}

\begin{figure}[ht]
  \includegraphics[width=14.cm]{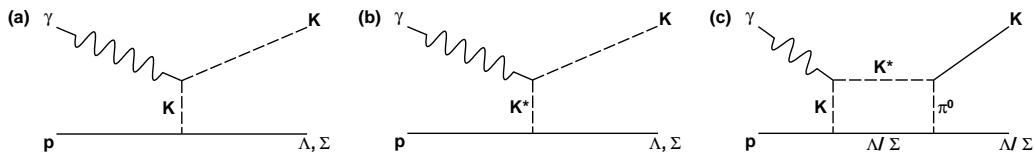}
  \caption{Contributions to kaon photoproduction (in addition to s-channel, u-channel and the contact term).  (a)~t-channel meson exchange, which only contributes in charged kaon photoproduction.  (b)~t-channel vector meson exchange, contributing in charged and neutral kaon photoproduction.  (c)~Subthreshold $K^*$ production and coupling to the kaon channel through $\pi^0$ rescattering.}\label{fig:feynman}
\end{figure}

The photoproduction of strange mesons is crucial in the search for ``missing'' nucleon resonances which have been predicted by constituent quark models but which have not been observed in non-strange photoproduction reactions~\cite{capstick94}.  To understand resonance structure, t-channel contributions must first be known before a partial wave analysis of s-channel contributions can be performed.  In the extensively studied channel, $\gamma(p,K^+)\Lambda$, there are large t-channel contributions from $K^+$ exchange (fig.\ref{fig:feynman}(a)).  As the photon cannot couple to $K^0$, this does not contribute in the lesser studied channel, $\gamma (p,K^+)\Sigma^0$, which is therefore a cleaner probe of s-channel contributions.

t-channel exchange of $K^*$ vector mesons, however, may still contribute (fig.~\ref{fig:feynman}(b)).  $K^0$ photoproduction is therefore an ideal tool to understand $K^*$-hyperon dynamics, and contributions such as fig.~\ref{fig:feynman}(c), which may exist as meson-hyperon dynamically generated states and may be evident close to the $K^*$ photoproduction threshold. 

\section{Experimental methods}

ELSA accelerated electrons to an energy of 3.2~GeV which were incident upon a thin radiator, producing bremstrahlung photons. The photon momentum was measured via the electron deflection in a magnetic dipole spectrometer before impinging upon a liquid hydrogen target at the centre of the Crystal Barrel.  For further details of the experimental setup see \cite{klein11,klein08}.

The reaction $\gamma(p,K^0)\Sigma^+$ was identified from the decays: $\Sigma^{+} \rightarrow p\pi^0$ and $K^0\rightarrow\pi^0\pi^0$.  Events were selected where six neutral particles were able to reconstruct the invariant mass of three $\pi^0$.  $\eta$ production was eliminated from the data set by rejecting events where the six neutral particle invariant mass was close to the $\eta$ mass.  The remaining events were subject to a least squares kinematic fit routine, rejecting events which yielded a confidence level less than 0.1.

\begin{figure}[ht]
  \includegraphics[width=10.cm]{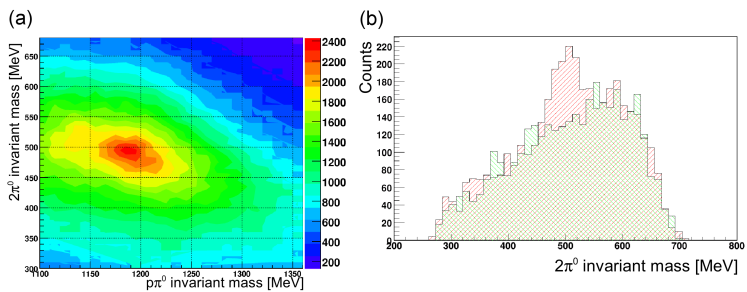}
  \caption{(a)~2$\pi^0$ invariant mass versus the $p\pi^0$ invariant mass, with a peak corresponding to the $K^0$ and $\Sigma^+$ masses.  (b)~2$\pi^0$ invariant mass after selecting events over the $\Sigma^+$ invariant mass range.  The (red) hatched region with postive gradient is experimental data, with a prominent peak at the $K^0$ mass. The (green) hatched region with negative gradient is simulated, uncorrelated 3$\pi^0$ background.  Taken from \cite{ralf}.}\label{fig:invmass}
\end{figure}

Fig.~\ref{fig:invmass}(a) is the 2$\pi^0$ invariant mass versus the proton-$\pi^0$ invariant mass.  Events were selected from the $p\pi^0$ invariant mass consistent with the $\Sigma^+$ mass (fig.~\ref{fig:invmass}(b)).  Background events were subtracted by comparison with simulated, uncorrelated $\gamma(p,p)3\pi^0$, the distribution of which matched the experimental data well.

The $4\pi$ acceptance and nearly flat detection detection efficiency over all energies and angles made the experiment ideal for cross section measurements.  For further details, see \cite{ralf}.

\section{Cross section measurements, interpretations and future plans}

\begin{figure}[ht]
  \includegraphics[width=12.5cm]{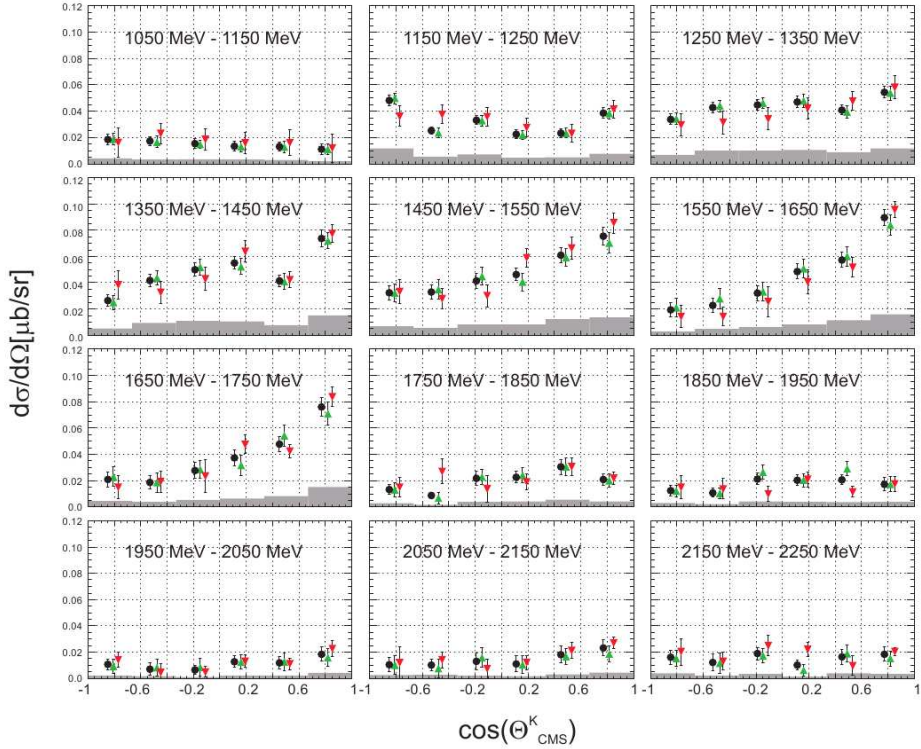}
  \caption{Differential cross section measurements for $\gamma(p,K^0)\Sigma^+$.  Data with one charged particle identified are (green) pointed up triangles, data with no charged particle identified are (red) pointed down triangles, (black) \index{}circles are all data combined.  Taken from~\cite{ralf}.}\label{fig:diffcrosssec}
\end{figure}

The $\gamma(p,K^0)\Sigma^+$ differential cross section (fig.~\ref{fig:diffcrosssec}) becomes increasingly forward peaked with energy, suggesting larger t-channel contributions.  Beyond a beam energy of 1750~MeV however, the cross section is flat.  At forward angles, the cross section drops by a factor of approximately four.

\begin{figure}[ht]
  \includegraphics[width=13.cm]{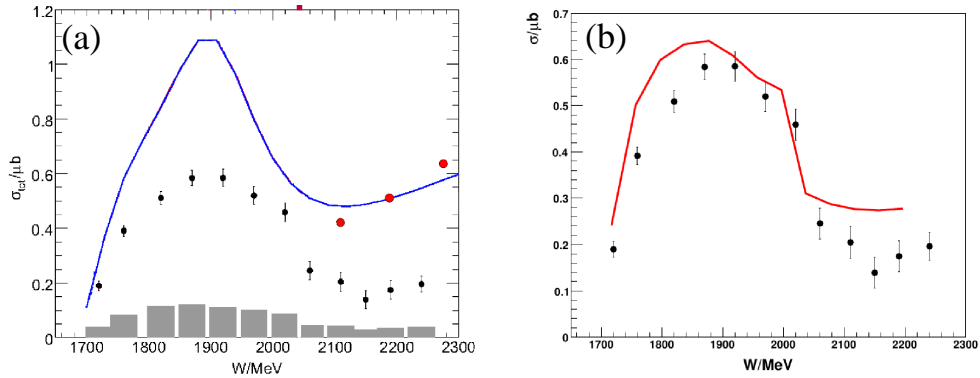}
  \caption{(a)~Cross section (small (black) circles) with additional data (larger (red) circles) of summed $\gamma (p,K^*)\Sigma^+$~\cite{nanova08} and $\gamma (p,K^0)\Sigma^+$ data and the kaon-MAID parameterisation~\cite{kaonmaid} (solid (blue) line).  (b)~Cross section with an adjusted Kaon-MAID~\cite{kaonmaid} parameterisation (see text).  Taken from~\cite{ralf}.}\label{fig:totcrosssec}
\end{figure}

This loss of strength at forward angles is apparent as a cusp-like structure in the total cross section (fig.~\ref{fig:totcrosssec}), occuring over the $K^*\Sigma^+$ and $K^*\Lambda$ thresholds.  It is speculated that there is large t-channel $K^*$ exchange in $\gamma(p,K^0)\Sigma^+$ below these energies.  Above threshold this no longer contributes, the $K^*$ is instead produced as a free particle.  Fig.~\ref{fig:totcrosssec}(b) shows an adjusted Kaon-MAID fit~\cite{kaonmaid}, where t-channel $K^*$ exchange has been ``switched off'' at the cusp-like structure energy, and the S$_{31}$(1900) couplings have been changed to 0.3 for both G$_1$ and G$_2$, giving a closer match to the data.  further evidence of $K^*$ subthreshold contribution can be seen in fig~\ref{fig:totcrosssec}(a), where above threshold, the $\gamma (p,K^*)\Sigma^+$ cross section~\cite{nanova08} has been added to the $\gamma(p,K^0)\Sigma^+$ cross section. This is in much better agreement with the existing Kaon-MAID parameterisation.

Just above threshold, $K^*$ couples strongly to a $K^0$ and $\pi^0$.  It is speculated that the production mechanism in fig.~\ref{fig:feynman}(c) feeds $K^0\Sigma^+$ photoproduction below the $K^*$ threshold.  The $\pi^0$ which is ejected from the $K^*$ is reabsorbed by the hyperon, and the $K^0$ is observed in the final state.  Above threshold, the $K^*$ is produced as a free particle and fig.~\ref{fig:feynman}(c) no longer contributes to the $\gamma(p,K^0)\Sigma^+$ cross section.

Recent chiral unitary appproaches have predicted dynamically generated states between vector mesons and baryons~\cite{oset10, oset11, sarkar10}, in particular, an isospin doublet with a mass of 1972~MeV, close to the $K^*$ production threshold.  A $K^*$-hyperon dynamically generated resonance occuring near the $K^*$ threshold would yield a $J^\pi$ of 1/2$^-$ or 3/2$^-$ (assuming the meson and hyperon to be relative s-wave).

The polarisation observable, $E$ (longitudinally polarised target, circularly polarised beam) provides a sensitive filter for the helicity structure of resonance contributions, and large changes of $E$ close to the $K^*$ threshold would provide an ideal test for dynamically generated states.  Moreover, $E$ would be sensitive to large changes in t-channel contributions, as pure t-channel contributions would give zero beam-target asymmetry.

Data has been taken with the CBELSA/TAPS experiment using a polarised butanol target to extract the polarisation observable, $E$, for $\gamma(p,K^0)\Sigma^+$.  Analysis is currently in progress.

The BGO Open Dipole (BGO-OD) experiment is in commision at the ELSA facility with first data expected to be taken in 2012.  BGO-OD consists of the BGO ball, a large acceptance calorimeter, designed for multiple photon detection with high energy and time resolution.  Forward angles are covered by a magnetic spectrometer with a series of tracking detectors, drift chambers and time of flight walls.  See \cite{klein11} for details.

The excellent forward angle acceptance and momentum reconstruction will be ideal for strangeness photoproduction, vector meson photoproduction and recoil polarisation measurements which will be complimentary to the results discussed here.

\begin{theacknowledgments}

On behalf of the CBELSA/TAPS and BGO-OD collaborations, supported by the German Science Foundation (DFG) in the framework of the SFB/TR-16.

\end{theacknowledgments}

\bibliographystyle{aipproc}   % if natbib is available

\end{document}